\documentclass[12pt]{iopart}
\usepackage{graphicx}
\newcommand{\cfigl}[3]{\begin{figure}[!hbtp]\centering
 \includegraphics[width=.5\textwidth]{#1}\caption{#2}\label{#3}\end{figure}}
\newcommand{\bfi }[1]{\mbox{\boldmath$#1$}}
\newcommand{\balpha}{\mbox{\boldmath$\alpha$}}
\newcommand{\bsigma}{\mbox{\boldmath$\sigma$}}
\usepackage{amsfonts}

\begin{document}

\title[The dynamical equation of the spinning electron]
{The dynamical equation of the spinning electron}

\author{Martin Rivas}
\address{Department of Theoretical Physics, University of the Basque Country,\\
Apdo 644, 48040 Bilbao, Spain}
\ead{wtpripem@lg.ehu.es}

\begin{abstract}
We obtain by invariance arguments the relativistic and non-relativistic invariant dynamical
equations of a classical model of a spinning electron. 
We apply the formalism to a particular
classical model which satisfies Dirac's equation when quantised.
It is shown that the dynamics can be described in terms of the evolution
of the point charge which satisfies a fourth order differential equation or, alternatively, 
as a system of second order differential equations 
by describing the evolution of both the center of 
mass and center of charge of the particle.
As an application of the found dynamical equations, the Coulomb interaction 
between two spinning electrons is considered. 
We find from the classical viewpoint that these spinning 
electrons can form bound states under suitable initial conditions. 
Since the classical Coulomb interaction of two spinless point electrons does not allow for
the existence of bound states, it is the spin structure that gives 
rise to new physical phenomena not described in the spinless case. Perhaps the 
paper may be interesting from the mathematical point of view but 
not from the point of view of physics.
\end{abstract}

\submitto{\JPA}
\maketitle

\section{Introduction}

Wigner defined a quantum elementary particle as a system
whose Hilbert space of states carries an Irreducible
Representation of the Poincar\'e group \cite{Wigner}. This definition is a group theoretical one.
This lead the author to look for a definition
of a classical elementary particle by group theoretical methods, relating its definition
to the kinematical group structure. A classical elementary particle was defined as 
a Lagrangian system whose kinematical space is a homogeneous space of 
the Poincar\'e group \cite{Rivas1,Rivasjm}. 
When quantising these classical models it is shown that the wave function of the system transforms
with a projective unitary irreducible representation of the kinematical group \cite{Rivas2}.
The different classical models of spinning particles
produced by this formalism are collected in the book \cite{Rivasl}. One of the models, 
which will be considered in this work, satisfies Dirac's equation when quantised.

The latest LEP experiments at CERN suggest that the electron charge is confined within a region
of radius $R_e<10^{-19}$m. Nevertheless, the quantum mechanical behaviour of the electron
appears at distances of the order of its Compton's wave length $\lambda_C=\hbar/mc\simeq 10^{-13}$m,
which is six orders of magnitude larger.
One possibility to reconcile these features in order to obtain a model of the electron 
is the assumption from the classical viewpoint
that the charge of the electron is just a point but at the same time this point is never at rest and it 
is affected by the so called {\em zitterbewegung} and therefore it is 
moving in a confined region of size $\lambda_C$. 

This is the basic structure of the spinning particle models obtained within the kinematical formalism
developed by the author \cite{Rivas1}-\cite{Rivasl} and also suggested by 
Dirac's analysis of the internal motion of the electron \cite{Dirac}. 
There, the charge of the particle is at a point ${\bfi r}$, but this point
is not the center of mass of the particle. In general we obtain that
the point charge satisfies a fourth order
differential equation which, as we shall see, is the most general differential equation
satisfied by any three-dimensional curve. 
The charge is moving around the center of mass in a kind
of harmonic or central motion. It is this motion of the charge that gives 
rise to the spin and dipole structure of the particle. 
In particular, the classical model that when quantised satisfies Dirac's equation
shows, for the center of mass observer, a charge moving at the speed of light in circles of 
radius $R=\hbar/2mc$ and contained in a plane orthogonal to the spin direction \cite{Rivas2}.
It is this classical model of electron we shall consider in the subsequent analysis and which is reviewed
in section \ref{sec:elec}.

In this article we shall find the group invariant dynamical equations of these classical systems.
The difference of the approach presented here with the previous published works is 
that the dynamical equations are obtained by group theoretical arguments without any appeal
to a Lagrangian formalism while there they were obtained by Lagrangian methods.
Neverteless the dynamical equations obtained are the same. 

The article is organised as follows: section 2 is a reminder that the most general 
differential equation satisfied by a curve in three-dimensional space is of a fourth order.
In section 3 we introduce the classical model of a spinning electron
that has been shown to satisfy Dirac's equation
when quantised. Section 4 states the general method for obtaining the group 
invariant differential equation satisfied by a point and
for any arbitrary kinematical group. 
This method is applied in sections 5 to 7 to the Galilei and Poincar\'e
groups to obtain the non-relativistic and relativistic invariant dynamical equations
of a spinless particle and of the spinning model.
Finally, as an application of the obtained dynamical equations, the analysis
of the Coulomb interaction between two spinning electrons is considered in section 8. 

One of the salient features is the classical prediction
of the possible existence of bound states for spinning electron-electron
interaction. If two electrons have their center of masses separated by a distance 
greater than Compton's wave length
they always repel each other. But if two electrons have their center of masses 
separated by a distance less than Compton's wave length
they can form bound states provided some conditions on their relative 
spin orientation and center of mass position and velocity are fulfilled.

\section{Frenet-Serret equations}

Let us remind that for an arbitrary three-dimensional curve ${\bfi r}(s)$ when 
expressed in parametric form in terms of the
arc length $s$, the three orthogonal unit 
vectors ${\bfi v}_i$, $i=1,2,3$ called respectively tangent, normal
and binormal, satisfy the so called Frenet-Serret differential equations:
\[
\begin{array}{ccccc}
\dot{\bfi v}_1(s)&=& &\kappa(s){\bfi v}_2(s)&\;\\
\dot{\bfi v}_2(s)&=&-\kappa(s){\bfi v}_1(s)& &+\tau(s){\bfi v}_3(s)\\
\dot{\bfi v}_3(s)&=& &-\tau(s){\bfi v}_2(s)&\;
\end{array}
\]
where $\kappa$ and $\tau$ are respectively the curvature and torsion.
Since the unit tangent vector is ${\bfi v}_1=\dot{\bfi r}\equiv{\bfi r}^{(1)}$, taking succesive derivatives 
it yields
 \begin{eqnarray*}
{\bfi r}^{(1)}&=&{\bfi v}_1,\\
{\bfi r}^{(2)}&=&\kappa{\bfi v}_2\\
{\bfi r}^{(3)}&=&\dot{\kappa}{\bfi v}_2+{\kappa}\dot{\bfi v}_2=-\kappa^2{\bfi v}_1+\dot{\kappa}{\bfi v}_2+\kappa\tau{\bfi v}_3,\\
{\bfi r}^{(4)}&=&-3\kappa\dot{\kappa}{\bfi v}_1+(\ddot{\kappa}-\kappa^3-\kappa\tau^2){\bfi v}_2+(2\dot{\kappa}\tau+\kappa\dot{\tau}){\bfi v}_3.
 \end{eqnarray*}
Then elimination of the ${\bfi v}_i$ between these equations implies 
that the most general curve
in three-dimensional space satisfies the fourth-order differential equation:
\[\fl
{\bfi r}^{(4)}-\left(\frac{2\dot{\kappa}}{\kappa}+\frac{\dot{\tau}}{\tau}\right){\bfi r}^{(3)}
+\left(\kappa^2+\tau^2+\frac{\dot{\kappa}\dot{\tau}}{\kappa\tau}+\frac{2\dot{\kappa}^2
-\kappa\ddot{\kappa}}{\kappa^2}\right){\bfi r}^{(2)}
+\kappa^2\left(\frac{\dot{\kappa}}{\kappa}-\frac{\dot{\tau}}{\tau}\right){\bfi r}^{(1)}=0
\]
All the coefficients in front of the derivatives ${\bfi r}^{(i)}$ can be expressed in terms
of the scalar products ${\bfi r}^{(i)}\cdot{\bfi r}^{(j)}$, $i,j=1,2,3$.
Let us mention that for helical motions there is a constant relationship $\kappa/\tau =$ constant,
and therefore the coefficient of ${\bfi r}^{(1)}$ vanishes.

\section{Spinning electron model}
\label{sec:elec}

We present here the main features of a spinning electron model obtained through
the general Lagrangian formalism. The Poincar\'e group can be parameterised in terms of the variables
$\{t,{\bfi r},{\bfi v},\balpha\}$, where $\balpha$ is dimensionless and represents the relative orientation
of inertial frames and the others have 
dimensions of time, length and velocity, respectively, 
and represent the corresponding parameters
for time and space translation and the relative velocity between observers. 
Their range
is $t\in\mathbb{R}$, ${\bfi r}\in\mathbb{R}^3$, ${\bfi v}\in\mathbb{R}^3$ 
but constrained to $v<c$ and $\balpha\in SO(3)$.

One of the important homogeneous spaces of the Poincar\'e group that defines the kinematical space
of a classical elementary particle \cite{Rivasjm} is spanned by the variables 
$\{t,{\bfi r},{\bfi v},\balpha\}$ with the same range as before but now ${\bfi v}$ restricted to $v=c$.
As variables of a kinematical space of a classical elementary system they are interpreted as
the time, position, velocity and orientation observables of the particle, respectively. 
The Lagrangian associated to this system will be a funtion of these variables and their 
next order time derivative.
Therefore the Lagrangian will also depend on the acceleration and angular velocity. It turns out that 
as far as the position ${\bfi r}$ is concerned dynamical equations will be of fourth order.
These are the equations we want to obtain in this work
but from a different and non-Lagrangian method. 
When analysed the invariance of any Lagrangian defined on this manifold 
under the Poincar\'e group, in particular 
under pure Lorentz transformations and rotations, we get respectively the following Noether constants 
of the motion which have the form $J^{\mu\nu}=-J^{\nu\mu}= x^\mu P^\nu-x^\nu P^\mu +S^{\mu\nu}$,
or its essential components $J^{0i}$ and $J^{ij}$ in three-vector notation \cite{Rivasjm}:
 \begin{equation}
{\bfi K}=\frac{1}{c^2}H{\bfi r}-{\bfi P}t-\frac{1}{c^2}{\bfi S}\times{\bfi v},
 \label{eq:K}
 \end{equation}
 \begin{equation}
{\bfi J}={\bfi r}\times{\bfi P}+{\bfi S}.
 \label{eq:J}
 \end{equation}
Observables $H$ and ${\bfi P}$ are the constant energy and linear momentum 
of the particle, respectively. The linear momentum is not lying along the velocity ${\bfi v}$,
and ${\bfi S}$ is the spin of the system. It satisfies the dynamical equation 
\[
\frac{d{\bfi S}}{dt}={\bfi P}\times{\bfi v},
\]
which is the classical equivalent of the dynamical equation satisfied by the 
Dirac spin observable.
For the center of mass observer it is a constant of the motion. This observer is defined
by the requirements ${\bfi P}=0$ and ${\bfi K}=0$. The first condition 
implies that the center of mass is at rest
and the second that is located at the origin of the frame. 
The energy of the particle in this frame
is $H=mc^2$, so that from (\ref{eq:K}) we get, 
\[
mc^2{\bfi r}={\bfi S}\times{\bfi v}.
\]
It turns out that point ${\bfi r}$, which does not represent the center of mass position, 
describes the motion depicted in Figure \ref{fig1}. It is the position of the charge.
It is orthogonal to the constant spin and also to the velocity as it corresponds
to a circular motion at the velocity $c$.
The radius and angular velocity of the 
internal classical motion of the charge are, 
respectively, $R={S}/{mc}$, and $\omega=c/R={mc^2}/{S}$.

\cfigl{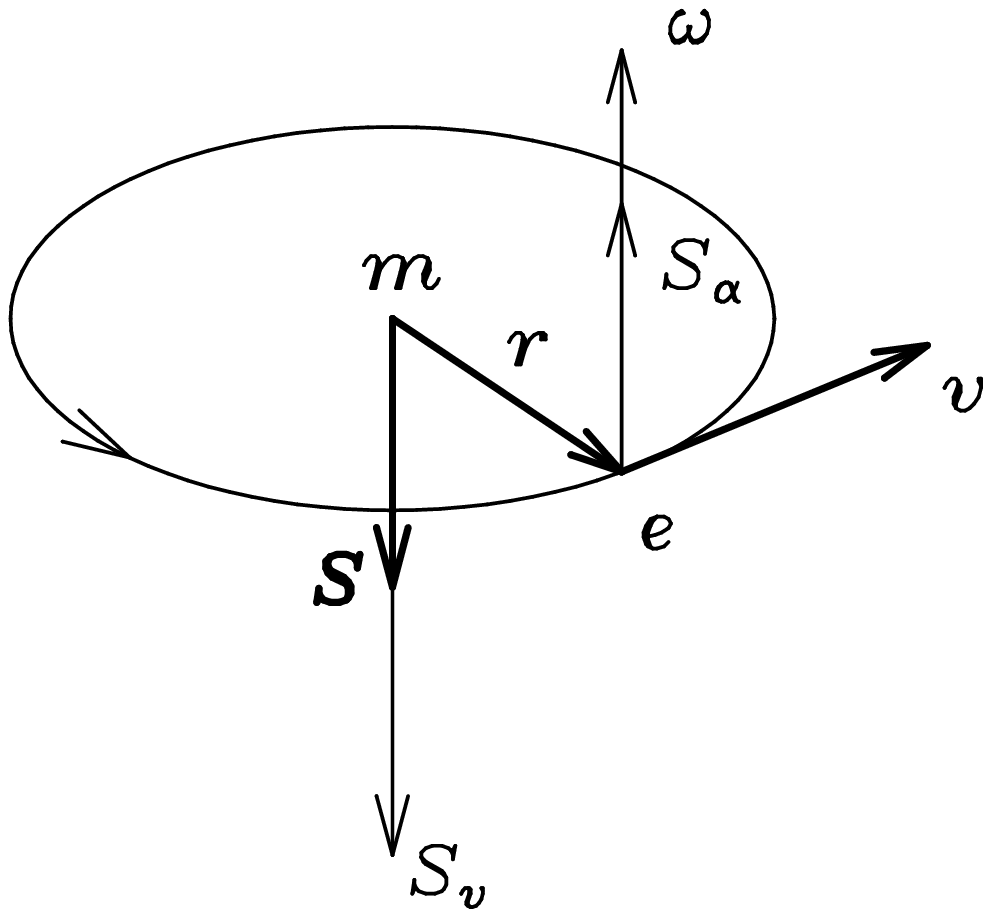}{Circular motion at the speed of light 
of the center of charge around the center of mass in the center of mass frame}{fig1}

If we take the time derivative of the constant ${\bfi K}$ of (\ref{eq:K}) 
and the scalar product with ${\bfi v}$ we get
\[
H-{\bfi P}\cdot{\bfi v}-{\bfi S}\cdot\left(\frac{d{\bfi v}}{dt}\times{\bfi v}\right)=0,
\]
where we have a linear relationship between $H$ and ${\bfi P}$.
Dirac equation is just the quantum mechanical expression of this Poincar\'e invariant
formula \cite{Rivas2}.

The spin ${\bfi S}={\bfi S}_v+{\bfi S}_\alpha$, 
has two parts: one ${\bfi S}_v$ related to the orbital motion of the charge and
another ${\bfi S}_\alpha$ due to the rotation of the particle and which 
is directly related to the angular velocity
as it corresponds to a spherically symmetric object. 
The positive energy particle has the total spin ${\bfi S}$ oriented 
in the same direction as the ${\bfi S}_v$ part, as shown in the figure. 
The orientation of the spin is the opposite for the negative energy particle, 
which corresponds to the time reversed motion.
When quantizing the system, the orbital component ${\bfi S}_v$ which is 
directly related to the magnetic moment,
quantizes with integer values while the rotational part ${\bfi S}_\alpha$ with half integer
values, so that for spin $1/2$ particles the total spin is half the value of the $S_v$ part.
When expressing the magnetic moment in terms of the total spin we get in this way a pure kinematical
interpretation of the $g=2$ gyromagnetic ratio \cite{Rivasetal}.

For the center of mass observer this system looks like a system of three degrees of freedom.
Two represent the $x$ and $y$ coordinates of the point and the third is the phase
of its rotational motion. However this phase is exactly equal to the phase of 
the orbital motion of the charge
and because the motion is of constant radius at the constant speed $c$ then only 
remains one independent degree of freedom.
Therefore the system is equivalent to a one-dimensional harmonic oscillator of 
frequency $\omega$. When quantizing the system 
the stationary states of the harmonic oscillator have the energy
\[
E_n=\left(n+\frac{1}{2}\right)\hbar\omega,\qquad n=0,1,2,\ldots
\]
But if the system is elementary then it has no excited states and in the C.M. frame 
it is reduced to the ground state of energy
\[
E_0=\frac{1}{2}\,\hbar\omega=mc^2,
\]
so that when compared with the classical result $\omega=mc^2/S$, 
implies that the constant classical 
parameter $S$ is necessarily $S={\hbar}/{2}$. 
The radius of the internal charge motion is 
half Compton's wave length.
It is this classical model of electron we shall analyse in subsequent sections, and our interest
is to obtain the dynamical equation satisfied by the point charge ${\bfi r}$ for any arbitrary
relativistic and nonrelativistic inertial observer.

To end this section and with the above model in mind let us collect the main results obtained by Dirac 
when he analysed the motion of the free electron  \cite{Dirac3}. Let 
point ${\bi r}$ be the position vector on which Dirac's spinor 
$\psi(t,{\bi r})$ is defined. When computing the velocity of point 
${\bi r}$, Dirac arrives at: 
\begin{enumerate}
\item{The velocity ${\bi v}=i/\hbar[H,{\bi r}]=c\balpha$, is expressed in terms of 
$\balpha$ matrices and writes, {\sl 
`$\ldots$ a measurement of a component of the velocity of a free 
electron is certain to lead to the result $\pm c$}. This conclusion is easily seen to hold also when there is a field present.'}

\item{ The linear momentum does not have the direction of this velocity 
${\bi v}$, but must be related to some average value of it: ${\ldots}$ 
{\sl `the $x_1$ component of the velocity, $c\alpha_1$, consists of 
two parts, a constant part $c^2p_1H^{-1}$, connected with the momentum 
by the classical relativistic formula, and an oscillatory part, whose 
frequency is at least $2mc^2/h$, ${\ldots}$'}. }

\item{ About the position ${\bi r}$: {\sl `The oscillatory part of $x_1$ is 
small, ${\ldots}$ , which is of order of magnitude $\hbar/mc$, 
${\ldots}$'}.}
\end{enumerate}
And when analysing, in his original 1928 paper \cite{Diracp} the 
interaction of the electron with an external electromagnetic field, 
after performing the square of Dirac's operator, he obtains two new 
interaction terms: 
 \begin{equation}
{e\hbar\over 2mc}{\bf\Sigma}\cdot{\bi B}+{ie\hbar\over 2mc}\balpha\cdot{\bi E},
 \label{eq:D8}
 \end{equation}
where the electron spin is written as ${\bi S}=\hbar{\bf\Sigma}/2$ and 
 \[ 
{\bf\Sigma}=\pmatrix{\bsigma&0\cr 0&\bsigma\cr},
 \] 
in terms of $\sigma$-Pauli matrices and ${\bi E}$ 
and ${\bi B}$ are the external electric and magnetic fields, 
respectively. He says, {\sl `The electron will therefore behave as 
though it has a magnetic moment $(e\hbar/2mc)\,{\bf\Sigma}$ and an 
electric moment $(ie\hbar/2mc)\,\balpha$. The magnetic moment
is just that assumed in the spinning electron model' }({\rm Pauli model}). `{\sl The 
electric moment, being a pure imaginary, we should not expect to 
appear in the model. It is doubtful whether the electric moment has any physical meaning.'} 

In the last sentence it is difficult to understand why Dirac, who did not reject the
negative energy solutions and therefore its consideration as the antiparticle states, 
disliked the existence of this electric dipole
which was obtained from his formalism on an equal footing as the magnetic dipole term.
In quantum electrodynamics, even in high energy processes, the complete Dirac Hamiltonian
contains both terms, perhaps in a rather involved way because the above expression 
is a first order expansion in the external fields considered as classical commuting fields. 
Properly speaking this electric dipole does not represent the existence of a particular 
positive and negative charge distribution for the electron. In the classical model, 
the negative charge of the electron
is at a single point but because this point is not the center of mass, there exists a nonvanishing
electric dipole moment with respect to the center of mass of value $e{\bfi r}$ in the center of mass frame. 
Its correspondance with the quantum Dirac electric moment is shown in \cite{Rivasl}. 
I think this is the observable Dirac disliked.
It is oscillating at very high frequency and it basically plays no role in low energy 
electron interactions because its average value vanishes, 
but it is important in high energy or in very close electron-electron interactions.

\section{The invariant dynamical equation}

Let us consider the trajectory ${\bfi r}(t)$, $t\in[t_1,t_2]$ followed by a point of a system for an 
arbitrary inertial observer $O$. Any other inertial observer $O'$ 
is related to the previous one by a transformation
of a kinematical group such that their relative space-time measurements of any space-time event are given by
\[
t'=T(t,{\bfi r}; g_1,\ldots,g_r),\quad {\bfi r}'={\bfi R}(t,{\bfi r}; g_1,\ldots,g_r),
\]
where the functions $T$ and ${\bfi R}$ define the action of the kinematical group $G$, 
of parameters $(g_1,\ldots,g_r)$, on space-time. Then the description of the trajectory of that point 
for observer $O'$ is obtained from
\[
t'(t)=T(t,{\bfi r}(t); g_1,\ldots,g_r),\quad {\bfi r}'(t)={\bfi R}(t,{\bfi r}(t); g_1,\ldots,g_r),\quad \forall t\in[t_1,t_2].
\]
If we eliminate $t$ as a function of $t'$ from the first equation and substitute into the second 
we shall get
\begin{equation}
{\bfi r}'(t')={\bfi r}'(t'; g_1,\ldots,g_r).
 \label{eq:rdet}
 \end{equation}
Since observer $O'$ is arbitrary, equation (\ref{eq:rdet}) represents the complete set of 
trajectories of the point for all inertial observers. 
Elimination of the $r$ group parameters among the function ${\bfi r}'(t')$
and their time derivatives will give us the differential equation satisfied by the trajectory of the point. 
This differential equation is invariant by construction because it is independent
of the group parameters and therefore independent of the inertial observer.
If $G$ is either the Galilei or Poincar\'e group
it is a ten-parameter group so that we have to work out in general up to the fourth derivative 
to obtain sufficient equations to eliminate the group parameters. 
Therefore the order of the differential equation is dictated by the
number of parameters and the structure of the kinematical group.

\section{The spinless particle}

Let us consider first the case of the spinless point particle.
In the non-relativistic case the relationship between inertial observers $O$ and $O'$
is given by the action of the Galilei group:
 \begin{equation}
t'=t+b,\quad {\bfi r}'=R(\balpha){\bfi r}+{\bfi v}t+{\bfi a}.
 \label{eq:trnon}
 \end{equation}
In the relativistic case we have that the Poincar\'e group action is given by
 \begin{eqnarray}
t'&=&\gamma\left(t+\frac{{\bfi v}\cdot R(\balpha){\bfi r}}{c^2}\right)+b,\label{eq:trt}\\
{\bfi r}'&=&R(\balpha){\bfi r}+\gamma{\bfi v}t+\frac{\gamma^2}{(1+\gamma)c^2}({\bfi v}\cdot 
R(\balpha){\bfi r})\,{\bfi v}+{\bfi a},\label{eq:trr}
 \end{eqnarray}
For the free spinless point particle it is possible
to find a particular observer, the center of mass observer $O^*$, 
such that the trajectory of the particle 
for this observer reduces to
\[
{\bfi r}^*(t^*)\equiv0,\quad\forall t^*\in[t_1^*,t_2^*].
\]
and therefore its trajectory for any other non-relativistic observer $O$ can be obtained from
 \begin{equation}
t(t^*)=t^*+b,\quad {\bfi r}(t^*)={\bfi v}t^*+{\bfi a}.
 \label{eq:non}
 \end{equation}
The trajectory of the point particle for the relativistic observer $O$ will be obtained from
 \begin{equation}
t(t^*)=\gamma t^*+b,\quad {\bfi r}(t^*)=\gamma{\bfi v}t^*+{\bfi a}.
 \label{eq:rel}
 \end{equation}
Elimination of $t^*$ in terms of $t$ from the first equation of both (\ref{eq:non}) and (\ref{eq:rel})
and substitution into the second yields
the trajectory of the point for an arbitrary observer, which in the relativistic and 
non-relativistic formalism reduces to
\[
{\bfi r}(t)=(t-b){\bfi v}+{\bfi a}.
\]
Elimination of group parameters ${\bfi v}$, $b$ and ${\bfi a}$ by taking succesive derivatives yields
the Galilei and Poincar\'e invariant dynamical equation of a free spinless point particle
\begin{equation}
\frac{d^2{\bfi r}}{dt^2}=0.
 \label{eq:rdos}
 \end{equation}

\section{The non-relativistic spinning electron}

We take spatial units such that the radius $R=1$, 
and time units such that $\omega=1$ and therefore the velocity $c=1$.
For the center of mass observer, the trajectory of the charge of the electron is
contained in the $XOY$ plane and it is expressed in 3-vector form as
\[
{\bfi r}^*(t^*)=\pmatrix{\cos t^*\cr \sin t^*\cr 0}.
\]
For the center of mass observer $O^*$ we get that
 \begin{equation}
\frac{d^2{\bfi r}^*(t^*)}{d{t^*}^2}=-{\bfi r}^*(t^*).
 \label{eq:rcm}
 \end{equation}
For any arbitrary inertial observer $O$ we get
 \begin{eqnarray*}
t(t^*;g)&=&t^*+b,\\
{\bfi r}(t^*;g)&=&R(\balpha){\bfi r}^*(t^*)+t^*{\bfi v}+{\bfi a}.
 \end{eqnarray*}
We shall represent the different time derivatives by
\[
{\bfi r}^{(k)}\equiv\frac{d^k{\bfi r}}{dt^k}=\frac{d}{d{t^*}}
\left(\frac{d^{k-1}{\bfi r}}{dt^{k-1}}\right)\frac{dt^*}{dt}.
\]
In this non-relativistic case $dt^*/dt=1$, then, after using (\ref{eq:rcm}) in some expressions 
we get the following derivatives
 \begin{eqnarray*}
{\bfi r}^{(1)}&=&R(\balpha)\frac{d{\bfi r}^*}{dt^*}+{\bfi v},\\
{\bfi r}^{(2)}&=&R(\balpha)\frac{d^2{\bfi r}^*}{d{t^*}^2}=-R(\balpha){\bfi r}^*,\\
{\bfi r}^{(3)}&=&-R(\balpha)\frac{d{\bfi r}^*}{dt^*},\\
{\bfi r}^{(4)}&=&-R(\balpha)\frac{d^2{\bfi r}^*}{d{t^*}^2}=R(\balpha){\bfi r}^*=-{\bfi r}^{(2)}.
 \end{eqnarray*}
Therefore, the differential equation satisfied by the position of the charge of a non-relativistic
electron and for any arbitrary inertial observer is
\begin{equation}
{\bfi r}^{(4)}+{\bfi r}^{(2)}=0.
 \label{eq:r4non}
 \end{equation}
We see that the motion is a helix because there is no ${\bi r}^{(1)}$ term.
\subsection{The center of mass}

The center of mass position of the electron is defined as
\begin{equation}
{\bfi q}={\bfi r}+{\bfi r}^{(2)}, 
 \label{eq:cm}
 \end{equation}
because it reduces to ${\bfi q}=0$ and ${\bfi q}^{(1)}=0$ for the center of mass observer, 
so that dynamical equations can be rewritten in terms of the position 
of the charge and the center of mass as
\begin{equation}
{\bfi q}^{(2)}=0,\quad {\bfi r}^{(2)}={\bfi q}-{\bfi r}.
 \label{eq:qyr}
 \end{equation}
Our fourth-order dynamical equation (\ref{eq:r4non}) can be split 
into two second order dynamical equations: A free
equation for the center of mass and a central harmonic 
motion of the charge position ${\bfi r}$ 
around the center of mass ${\bfi q}$ of angular frequency 1 in these natural units.

\subsection{Interaction with some external field}

The free dynamical equation ${\bfi q}^{(2)}=0$ is equivalent to $d{\bfi P}/dt=0$, 
where ${\bfi P}=m{\bfi q}^{(1)}$
is the linear momentum of the system. Then our free equations should be replaced in the 
case of an interaction with an external electromagnetic field by
\begin{equation}
m{\bfi q}^{(2)}=e[{\bfi E}+{\bfi r}^{(1)}\times{\bfi B}],\quad {\bfi r}^{(2)}={\bfi q}-{\bfi r},
\label{eq:Galilei}
\end{equation}
where in the Lorentz force the fields are defined at point ${\bfi r}$ and 
it is the velocity of the charge that gives 
rise to the magnetic force term, while the second equation is left unchanged 
since it corresponds to the center of mass definition. 

These equations are also obtained in the Lagrangian approach \cite{Rivasl} by assuming a minimal coupling
interaction and where we get
 \begin{equation}
{\bfi r}^{(4)}+{\bfi r}^{(2)}=\frac{e}{m}[{\bfi E}(t,{\bfi r})+{\bfi r}^{(1)}\times{\bfi B}(t,{\bfi r})],
 \label{eq:Lag}
 \end{equation}
which reduce to (\ref{eq:Galilei}) after the center of mass definition (\ref{eq:cm}). 

In order to determine the evolution of the system, initial conditions ${\bfi r}(0)$, ${\bfi r}^{(1)}(0)$,
${\bfi r}^{(2)}(0)$ and ${\bfi r}^{(3)}(0)$, i.e., the position of point ${\bfi r}$ and its derivatives
up to order 3 evaluated at time $t=0$ must be given. Alternatively, if we consider that our fourth
order differential equation (\ref{eq:Lag}) as the set of two second order differential equations
(\ref{eq:Galilei}), then we have to fix as initial conditions ${\bfi r}(0)$ and ${\bfi r}^{(1)}(0)$
as before and ${\bfi q}(0)={\bfi r}(0)+{\bfi r}^{(2)}(0)$ 
and ${\bfi q}^{(1)}(0)={\bfi r}^{(1)}(0)+{\bfi r}^{(3)}(0)$, compatible with (\ref{eq:cm}), i.e.,
the position and velocity of both points the center of mass and center of charge. The advantage
of this method is that we shall be able to analyse the evolution of a 2-electron system in 
section \ref{sec:2elec} in terms of the center of mass initial position and velocity.

\section{The relativistic spinning electron}

Let us assume the same electron model in the relativistic case. Since the charge is moving at the speed
of light for the center of mass observer $O^*$ it is moving 
at this speed for every other inertial observer $O$. 
Now, the relationship of space-time measurements between the center of mass observer and 
any arbitrary inertial observer is given by:
 \begin{eqnarray*}
t(t^*;g)&=&
\gamma\left(t^*+{\bfi v}\cdot R(\balpha){\bfi r}^*(t^*)\right)+b,\\
{\bfi r}(t^*;g)&=&
R(\balpha){\bfi r}^*(t^*)+\gamma{\bfi v}t^*+\frac{\gamma^2}{1+\gamma}
\left({\bfi v}\cdot R(\balpha){\bfi r}^*(t^*)\right){\bfi v}+{\bfi a}.
 \end{eqnarray*}
With the shorthand notation for the following expressions:
\[
{\bfi K}(t^*)=R(\balpha){\bfi r}^*(t^*),
\quad {\bfi V}(t^*)=R(\balpha)\frac{d{\bfi r}^*(t^*)}{dt^*}=
\frac{d{\bfi K}}{dt^*},\quad
\frac{d{\bfi V}}{dt^*}=-{\bfi K},
\]
\[
B(t^*)={\bfi v}\cdot{\bfi K},\quad
A(t^*)={\bfi v}\cdot{\bfi V}=\frac{dB}{dt^*},
\quad\frac{dA}{dt^*}=-B 
\]
we obtain
\begin{eqnarray}
{\bfi r}^{(1)}&=&\frac{1}{\gamma(1+A)}\left({\bfi V}+\frac{\gamma}{1+\gamma}
(1+\gamma+\gamma A){\bfi v}\right),\label{eq:r1}\\
{\bfi r}^{(2)}&=&\frac{1}{\gamma^2(1+A)^3}\left(-(1+A){\bfi K}+B{\bfi V}
+\frac{\gamma}{1+\gamma}\,B{\bfi v}\right),\label{eq:r2}
\end{eqnarray}
\[
{\bfi r}^{(3)}=\frac{1}{\gamma^3(1+A)^5}\left(-3B(1+A){\bfi K}-(1+A-3B^2){\bfi V}+\right.\]
\begin{equation}
\qquad\left.\frac{\gamma}{1+\gamma}\,(A(1+A)+3B^2){\bfi v}\right)\label{eq:r3}
\end{equation}
\[
{\bfi r}^{(4)}=\frac{1}{\gamma^4(1+A)^7}\left((1+A)(1-2A-3A^2-15B^2){\bfi K}-\right.\]
\[
\qquad B(7+4A-3A^2-15B^2){\bfi V}-\]
\begin{equation}
\qquad\left.\frac{\gamma}{1+\gamma}\,(1-8A-9A^2-15B^2)B{\bfi v}\right).\label{eq:r4}
\end{equation}
From this we get
 \begin{eqnarray}
\left({\bfi r}^{(1)}\cdot{\bfi r}^{(1)}\right)^2&=&1,\quad\left({\bfi r}^{(1)}\cdot{\bfi r}^{(2)}\right)=0,
\label{eq:inv1}\\
 \left({\bfi r}^{(2)}\cdot{\bfi r}^{(2)}\right)&=&-\left({\bfi r}^{(1)}\cdot{\bfi r}^{(3)}\right)=\frac{1}{\gamma^4(1+A)^4},
\label{eq:inv2}\\
\left({\bfi r}^{(2)}\cdot{\bfi r}^{(3)}\right)&=&-\frac{1}{3}\left({\bfi r}^{(1)}\cdot{\bfi r}^{(4)}\right)=\frac{2B}{\gamma^5(1+A)^6},
\label{eq:inv3}\\
\left({\bfi r}^{(3)}\cdot{\bfi r}^{(3)}\right)&=&\frac{1}{\gamma^6(1+A)^8}\left(1-A^2+3B^2\right),
\label{eq:inv4}\\
\left({\bfi r}^{(2)}\cdot{\bfi r}^{(4)}\right)&=&\frac{1}{\gamma^6(1+A)^8}\left(-1+2A+3A^2+9B^2\right),
\label{eq:inv5}\\
\left({\bfi r}^{(3)}\cdot{\bfi r}^{(4)}\right)&=&\frac{1}{\gamma^7(1+A)^{10}}\left(1+A+3B^2\right)4B.
\label{eq:inv6}
 \end{eqnarray}
From equations (\ref{eq:inv2})-(\ref{eq:inv4}) we can express the magnitudes $A$, $B$ and $\gamma$ in terms of these
scalar products between the different time derivatives $({\bfi r}^{(i)}\cdot{\bfi r}^{(j)})$. The constraint
that the velocity is 1 implies that all these and further scalar products for higher derivatives can be expressed
in terms of only three of them.
If the three equations (\ref{eq:r1})-(\ref{eq:r3})
are solved in terms of the unknowns ${\bfi v}$, ${\bfi V}$ and ${\bfi K}$ and substituded into (\ref{eq:r4}),
we obtain the differential equation satisfied by the charge position
\[
{\bfi r}^{(4)}-\frac{3({\bfi r}^{(2)}\cdot{\bfi r}^{(3)})}{({\bfi r}^{(2)}\cdot{\bfi r}^{(2)})}\,{\bfi r}^{(3)}+\]
\begin{equation}
\qquad\left(\frac{2({\bfi r}^{(3)}\cdot{\bfi r}^{(3)})}{({\bfi r}^{(2)}\cdot{\bfi r}^{(2)})}-
\frac{3({\bfi r}^{(2)}\cdot{\bfi r}^{(3)})^2}{4({\bfi r}^{(2)}\cdot{\bfi r}^{(2)})^2}-({\bfi r}^{(2)}\cdot{\bfi r}^{(2)})^{1/2}\right){\bfi r}^{(2)}=0.\label{eq:elecbuena}
\end{equation}
It is a fourth order ordinary differential equation which contains as solutions 
motions at the speed of light. In fact, if $({\bfi r}^{(1)}\cdot{\bfi r}^{(1)})=1$, then by derivation we have $({\bfi r}^{(1)}\cdot{\bfi r}^{(2)})=0$
and the next derivative leads to  $({\bfi r}^{(2)}\cdot{\bfi r}^{(2)})+({\bfi r}^{(1)}\cdot{\bfi r}^{(3)})=0$. If we take this into account
and make the scalar product of (\ref{eq:elecbuena}) with ${\bfi r}^{(1)}$, we get $({\bfi r}^{(1)}\cdot{\bfi r}^{(4)})+3({\bfi r}^{(2)}\cdot{\bfi r}^{(3)})=0$,
which is another relationship between the derivatives as a consequence of $|{\bfi r}^{(1)}|=1$.
It corresponds to a helical motion since the term in the first derivative ${\bfi r}^{(1)}$ is lacking.

\subsection{The center of mass}

The center of mass position is defined by
\begin{equation}
{\bfi q}={\bfi r}+\frac{2({\bfi r}^{(2)}\cdot{\bfi r}^{(2)})\,{\bfi r}^{(2)}}{({\bfi r}^{(2)}\cdot{\bfi r}^{(2)})^{3/2}+({\bfi r}^{(3)}\cdot{\bfi r}^{(3)})-
\frac{\displaystyle{3({\bfi r}^{(2)}\cdot{\bfi r}^{(3)})^2}}{\displaystyle{4({\bfi r}^{(2)}\cdot{\bfi r}^{(2)})}}}.
 \label{eq:cmq}
 \end{equation}
We can check that both ${\bfi q}$ and ${\bfi q}^{(1)}$ vanish for the center of mass observer. 
Then, the fourth order dynamical equation for the position of the 
charge can also be rewritten here
as a system of two second order differential equations for the positions ${\bfi q}$ and ${\bfi r}$
\begin{equation}
{\bfi q}^{(2)}=0,\quad {\bfi r}^{(2)}=
\frac{1-{\bfi q}^{(1)}\cdot{\bfi r}^{(1)}}{({\bfi q}-{\bfi r})^2}\left({\bfi q}-{\bfi r}\right),
\label{eq:q2r2}\end{equation}
a free motion for the center of mass and a kind of central motion for the 
charge around the center of mass.

For the non-relativistic electron we get in the low velocity case ${\bfi q}^{(1)}\to0$
and $|{\bfi q}-{\bfi r}|=1$, the equations of the Galilei case 
\begin{equation}
{\bfi q}^{(2)}=0,\quad {\bfi r}^{(2)}=
{\bfi q}-{\bfi r}.
\label{eq:q321}\end{equation}
a free motion for the center of mass and a harmonic motion
around ${\bfi q}$ for the position of the charge. 

\subsection{Interaction with some external field}

The free equation for the center of mass motion ${\bfi q}^{(2)}=0$, represents the conservation
of the linear momentum $d{\bfi P}/dt=0$. But the linear momentum is 
written in terms of the center of mass velocity as
${\bfi P}=m\gamma(q^{(1)}){\bfi q}^{(1)}$, so that the free dynamical 
equations (\ref{eq:q2r2}) in the presence
of an external field should be replaced by
\begin{equation}
{\bfi P}^{(1)}={\bfi F},\quad {\bfi r}^{(2)}=
\frac{1-{\bfi q}^{(1)}\cdot{\bfi r}^{(1)}}{({\bfi q}-{\bfi r})^2}\left({\bfi q}-{\bfi r}\right),
\label{eq:Pr2}\end{equation}
where ${\bfi F}$ is the external force and the second equation is left unchanged because we consider, 
even with interaction, the same definition of the center of mass position. 
\[
\frac{d{\bfi P}}{dt}=m\gamma(q^{(1)}){\bfi q}^{(2)}+m\gamma(q^{(1)})^3({\bfi q}^{(1)}\cdot{\bfi q}^{(2)}){\bfi q}^{(1)}
\]
we get
\[
m\gamma(q^{(1)})^3({\bfi q}^{(1)}\cdot{\bfi q}^{(2)})={\bfi F}\cdot{\bfi q}^{(1)}
\]
and by leaving the highest derivative ${\bfi q}^{(2)}$ on the left hand side we finally get the differential
equations which describe the evolution of a relativistic spinning electron in the presence of an 
external electromagnetic field:
\begin{eqnarray}
m{\bfi q}^{(2)}&=&\frac{e}{\gamma(q^{(1)})}\left[{\bfi E}+{\bfi r}^{(1)}\times{\bfi B}-{\bfi q}^{(1)}\left(\left[{\bfi E}
+{\bfi r}^{(1)}\times{\bfi B}\right]\cdot{\bfi q}^{(1)}\right)\right],\\
{\bfi r}^{(2)}&=&\frac{1-{\bfi q}^{(1)}\cdot{\bfi r}^{(1)}}{({\bfi q}-{\bfi r})^2}\left({\bfi q}-{\bfi r}\right).
 \end{eqnarray}

\section{Two electron system}
\label{sec:2elec}

If we have the relativistic and nonrelativistic differential equations satisfied by the charge
of the spinning electrons
we can analyse as an example, 
the interaction among them by assuming a Coulomb interaction between their charges.
In this way we have a system of differential equations of the form (\ref{eq:Galilei}) for each particle.
For instance, the external field acting on charge $e_1$ is replaced by the Coulomb 
field created by the other charge $e_2$
at the position of $e_1$, and simmilarly for the other particle.
The integration is performed numerically by means of the numerical integration program {\it Dynamics Solver}
\cite{JMA}.

\vspace{-1cm}
\cfigl{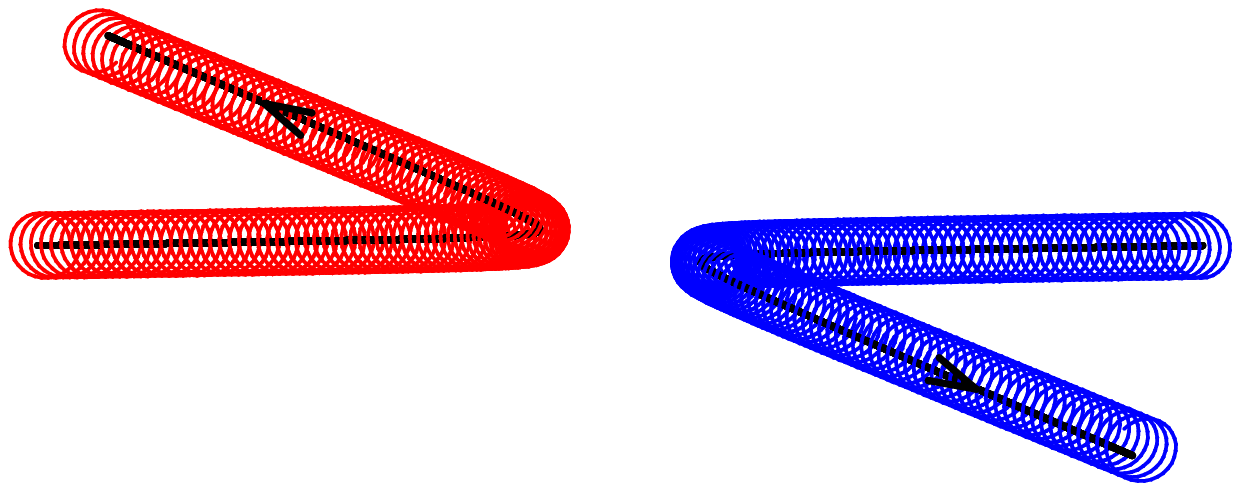}{Scattering of two spinning electrons with the spins parallel, in their center of mass frame.
It is also depicted the scattering of two spinless electrons with the same energy and linear momentum.}
{fig2}

In Figure \ref{fig2} we represent the scattering of two spinning electrons 
analysed in their center of mass frame. We send the particles with their spins parallel and 
with a nonvanishing impact parameter. In addition to the curly motion of their charges we can also 
depict the trajectories of their center of masses. If we compare 
this motion with the Coulomb interaction
of two spinless electrons coming from the same initial position and with the same 
velocity as the center of mass 
of the spinning electrons we obtain the solid trajectory marked with an arrow. 
Basically this corresponds to the
trajectory of the center of mass of each spinning particle provided the two particles 
do not approach each other below Compton's wave length. 
This can be understood because the average position of the center of charge
of each particle aproximately coincides with its center of mass and as far as they do not approach
each other too much the average Coulomb force is the same. The difference comes out when we consider
a very deep interaction or very close initial positions.

In Figure \ref{fig3} we represent the initial positions for a pair of particles
with the spins parallel. The initial separation $a$ of their center of masses is 
a distance below Compton's wave length. We also consider that initially the center of mass of each
particle is moving with a velocity $v$ as depicted.
That spins are parallel is reflected
by the fact that the internal motions of the charges, represented by the oriented circles that surround
the corresponding center of mass, have the same orientation.
It must be remarked that the charge motion around its center of mass can be characterised
by a phase. The phases of each particle are chosen 
opposite to each other.
We also represent the repulsive Coulomb force $F$ computed in terms of the separation 
of the charges. 
This interacting force $F$ has also been attached to the corresponding center of mass,
so that the net force acting on point $m_2$ is directed towards point $m_1$, and conversely. 
We thus see there that a repulsive 
force between the charges represents an atractive force between their center of masses when located
at such a short distance.

\cfigl{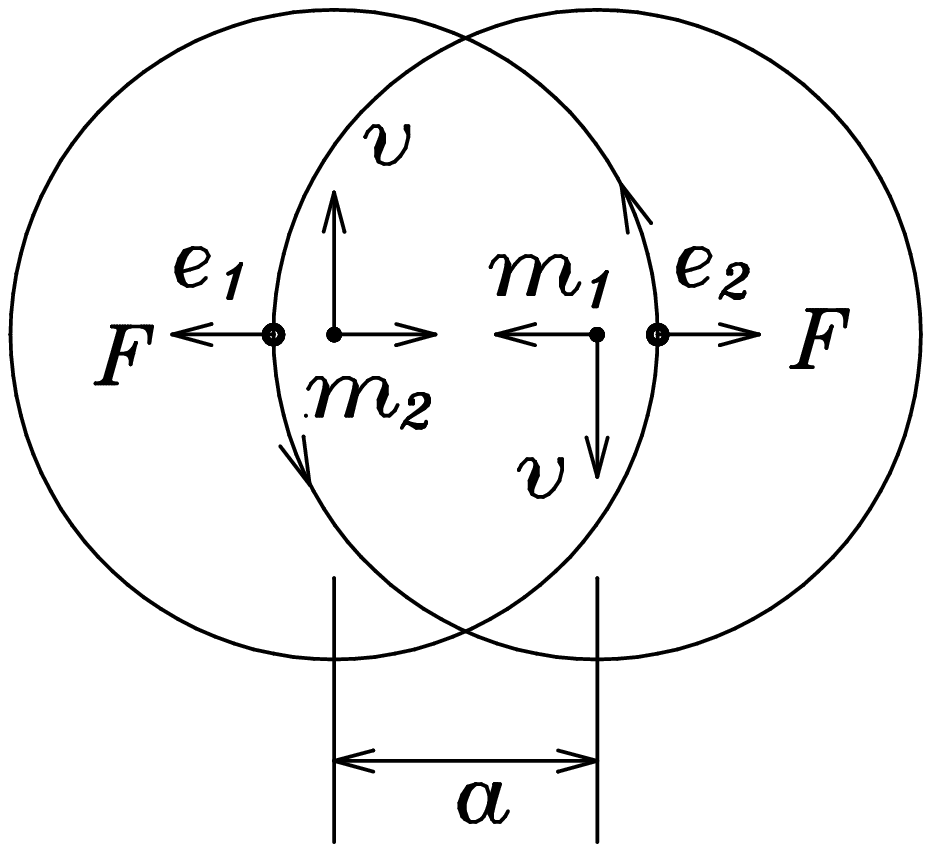}{Initial position and velocity of the center of mass and charges for a bound motion 
of a two-electron system with parallel spins. The circles would correspond to the 
trajectories of the charges if considered free. The interacting Coulomb force $F$ is computed
in terms of the separation distance between the charges.}{fig3}

In Figure \ref{fig4} we depict the evolution of the charges and masses 
of this two-electron system for $a=0.4\lambda_C$ and $v=0.004c$ during a short time interval. 
Figure \ref{fig5} represents only the motions of the center of masses
of both particles for a longer time. 
It shows that the center of mass of each particle 
remains in a bound region.

\cfigl{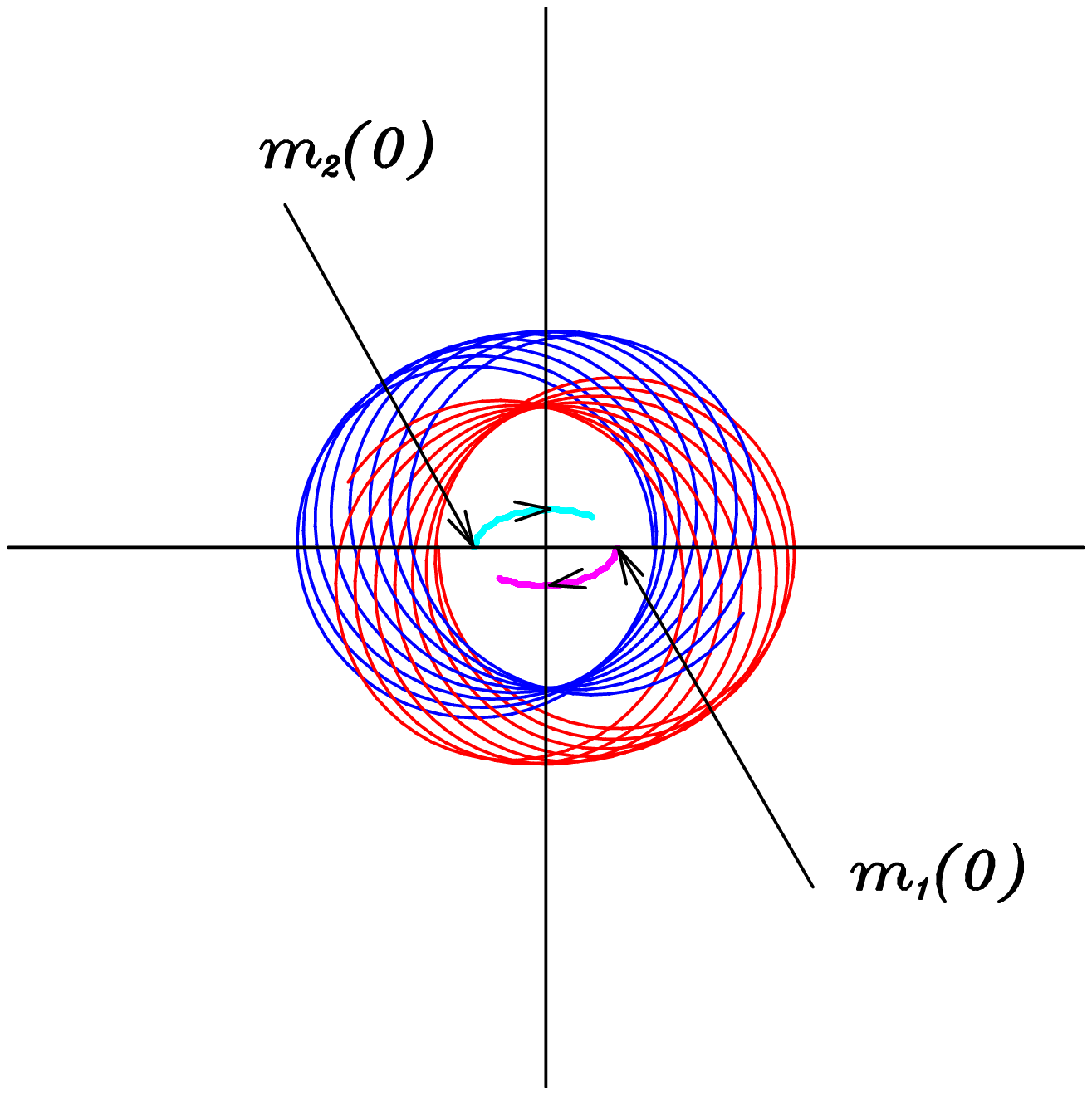}{Bound motion of two electrons with parallel spins during a short period of time}{fig4}

The evolution of the charges is not shown in this figure because it blurs the picture
but it can be inferred from the previous figure.
We have found bound motions at least for the range $0\le a\le 0.8\lambda_C$ and velocity
$0\le v\le 0.01c$. We can also obtain similar bound motions if the initial velocity $v$ 
has a component along the $OX$ axis. 
The bound motion is also obtained for different initial charge positions
as the ones depicted in Figure \ref{fig3}. This range for the relative phase 
depends on $a$ and $v$ but in general the bound motion is more likely if the 
initial phases of the charges are opposite to each other.

We thus see that if the separation 
between the center of mass and center of charge of a particle ({\it zitterbewegung})
is responsible for its spin structure as has been shown in 
the formalism developped by the author then this atractive 
effect and also a spin polarised tunneling effect can be easily interpreted
\cite{Rivas3}.

\cfigl{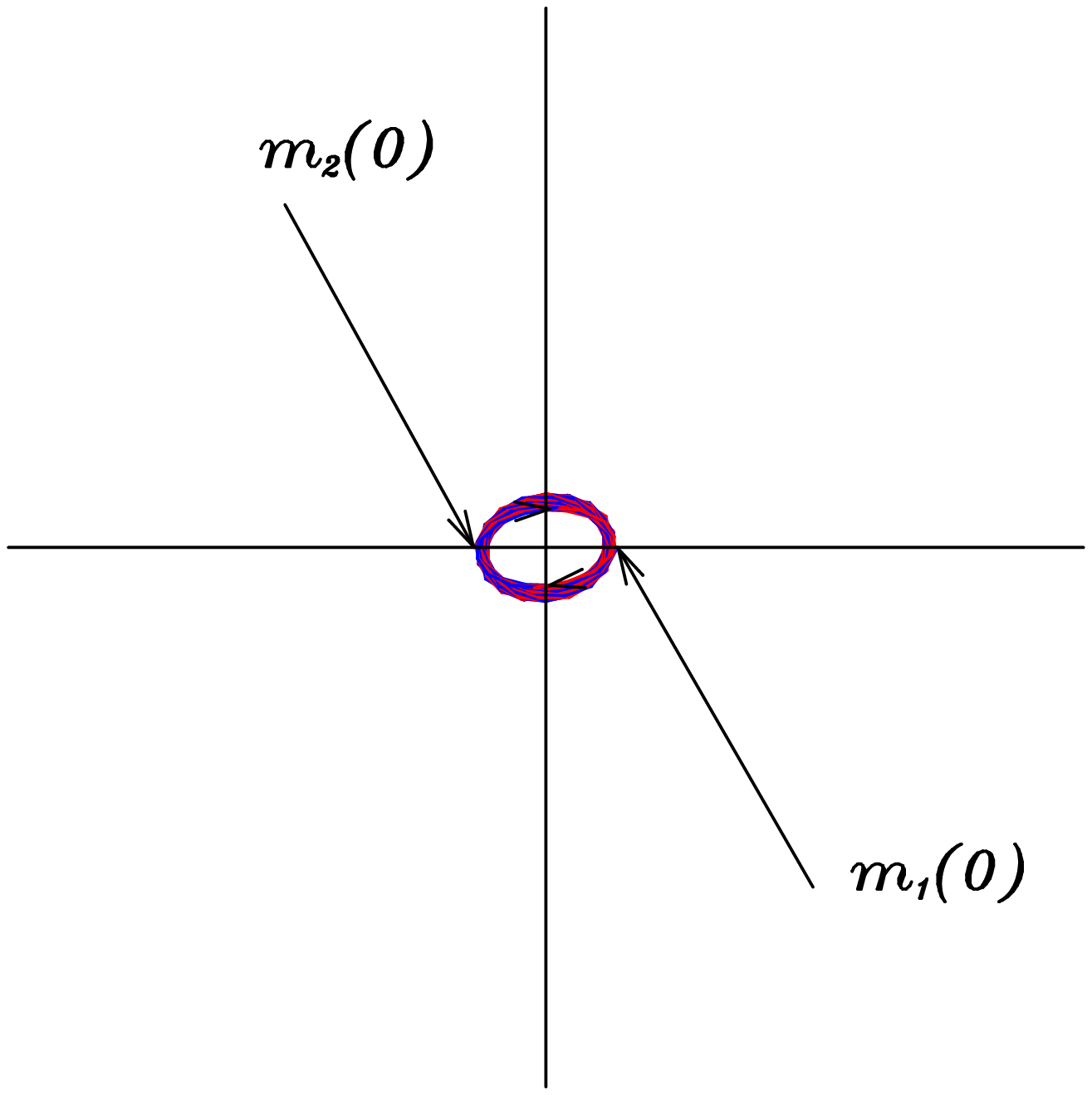}{Evolution of the center of mass of both particles for a larger time}{fig5}

A bound motion for classical spinless electrons is not possible. 
We can conclude that one of the salient features of this example
is the existence from the classical 
viewpoint of bound states for spinning electron-electron interaction. 
It is the spin structure which contributes to the prediction of new physical phenomena.
If two electrons have their center of masses separated by 
a distance greater than Compton's wave length
they always repel each other as in the spinless case.
But if two electrons have their center of masses separated by a distance less 
than Compton's wave length
they can form from the classical viewpoint bound states provided some 
initial conditions on their relative 
initial spin orientation, position of the charges and 
center of mass velocity are fulfilled. 

The example analysed gives just a classical prediction, not a quantum one, 
associated to a model that satisfies
Dirac's equation when quantised. The possible 
quantum mechanical bound states if they exist, must be obtained from the 
corresponding analysis of two interacting quantum Dirac particles, bearing in mind 
that the classical bound states are not forbiden from the classical viewpoint.
Bound states for a hydrogen atom can exist from the classical viewpoint for any
negative energy and arbitrary angular momentum. 
It is the quantum analysis of the atom that gives the correct
answer to the allowed bound states. 
The quantum mechanical analysis of a two electron system is left to a subsequent paper.

\ack{I thank my colleague J M Aguirregabiria for the use of his excellent
Dynamics Solver program which I strongly recommend even as a teaching tool. 
This work was supported by The University of the Basque Country under the project
UPV/EHU 172.310 EB 150/98 and General Research Grant UPV 172.310-G02/99.}

\end{document}